\documentclass[final]{aipproc}
\layoutstyle{6x9}
\usepackage{graphicx}
\usepackage{epsf}
\usepackage{wrapfig}
\usepackage{sublabel}
\usepackage{epsfig}
\usepackage{amsmath}

 \def\gsim{\mathrel{\rlap{\lower4pt\hbox{\hskip1pt$\sim$}}
 \raise1pt\hbox{$>$}}}

 \newcommand\ra{\rangle}
 \newcommand\beq{\begin{equation}}
 
 \newcommand\eeq{\end{equation}}
 \newcommand\beqn{\begin{eqnarray}}
 \newcommand\eeqn{\end{eqnarray}}

\def\GeV{\,\mbox{GeV}}

\def\lsim{\mathrel{\rlap{\lower4pt\hbox{\hskip1pt$\sim$}}
    \raise1pt\hbox{$<$}}}         
\def\gsim{\mathrel{\rlap{\lower4pt\hbox{\hskip1pt$\sim$}}
    \raise1pt\hbox{$>$}}}         

\def\Re{\,{\rm Re}\,}
\def\Im{\,{\rm Im}\,}

\def\GeV{\,\mbox{GeV}}

\newcommand{\be}{\begin{equation}}
\newcommand{\ee}{\end{equation}}

\def\b0{{\mbox{\boldmath$0$}}}

\def\b0{{\mbox{\boldmath$0$}}}

\def \b #1{ {\bf #1}}

\def \b #1{ {\bf #1}}

\def\beqy{\begin{eqnarray}}
\def\eeqy{\end{eqnarray}}

\begin{document}
\title{Leading neutrons from polarized pp collisions}


\author{B.Z.~Kopeliovich}{
  address={Departamento de F\'{\i}sica y Centro de Estudios
Subat\'omicos,\\
Universidad T\'ecnica Federico Santa Mar\'{\i}a,
Casilla 110-V,
Valpara\'\i so, Chile},
altaddress={Institut f\"ur Theoretische Physik der Universit\"at,
Philosophenweg 19, 69120
Heidelberg, Germany}}

\author{I.K.~Potashnikova}{
  address={Departamento de F\'{\i}sica y Centro de Estudios
Subat\'omicos,\\
Universidad T\'ecnica Federico Santa Mar\'{\i}a,
Casilla 110-V,
Valpara\'\i so, Chile}
}

\author{Ivan~Schmidt}{
  address={Departamento de F\'{\i}sica y Centro de Estudios
Subat\'omicos,\\
Universidad T\'ecnica Federico Santa Mar\'{\i}a,
Casilla 110-V,
Valpara\'\i so, Chile}
}
\author{J.~Soffer}{
  address={Department of Physics, Temple University, Philadelphia, PA 19122-6082, 
USA}
}

\begin{abstract}
 We calculate the cross section and single-spin azimuthal
asymmetry, $A_n(t)$ for inclusive neutron production in $pp$
collisions at forward rapidities relative to the polarized proton.
Absorptive corrections to the pion pole generate a relative phase between
the spin-flip and non-flip amplitudes, which leads to an appreciable spin asymmetry.
However, the asymmetry observed recently in the PHENIX experiment at RHIC
at very small $|t|\sim 0.01\GeV^2$ cannot be explained by this mechanism.
\end{abstract}

\maketitle

\section{Introduction}

\begin{figure}[htb]
\centerline{
 {\includegraphics[height=6cm]{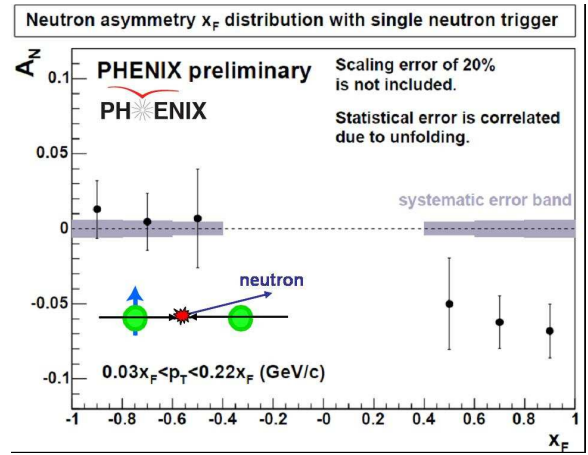}}
 }
\caption{Single-spin asymmetry $A_n$ in reaction $pp\to nX$ measured at $\sqrt{s}=200\GeV$ \cite{spin2006} \label{rhic1}}
 \end{figure}

 We are considering the reaction of inclusive neutron production $pp\to Xn$ at large 
Feynman $x_F$ as was measured recently by the PHENIX experiment at RHIC \cite{phenix} at $\sqrt{s} = 200\GeV$. The preliminary data \cite{spin2006} are shown in Fig.~\ref{rhic1}.
The measurements were done with a transversely polarized proton beam and the neutron was detected in the very near forward (or backward) 
direction relative to the polarized beam with a transverse momentum $p_T$ of the 
order of 100 MeV. An appreciable single spin asymmetry was found for forward 
neutrons, for events with neutron momenta larger than $30\GeV$ $A_N=(-9.0 \pm 0.6 \pm 0.9)\% $ with an additional scale uncertainty $\left(1.0 {\footnotesize
\begin{array}{c}+0.47\\-0.24 \end{array}}\right)$. Data 
also show no dramatic variation with $x_F$. According to the Feynman 
scaling, one should not expect any strong energy dependence of $A_N(z)$.  

At the same time, neutrons produced with $x_F<0$ demonstrate small asymmetry, 
consistent with  zero. This fact is explained by the so called 
Abarbanel-Gross theorem \cite{ag} which predicts zero asymmetry for particles
produced in the fragmentation region of an unpolarized beam.
 This theorem was 
proven within the approximation of triple-Regge poles. Regge cuts breakdown this statement, but the corrections calculated in \cite{klz} turned out to be rather small,
less than $1\%$.

The cross section of of neutron production in $pp$ collisions  was calculated recently in \cite{kpss} where it was found that previous calculations \cite{3n,ap}, although agreed with the ISR data, grossly underestimated the magnitude of absorptive corrections. New measurements done recently show that the normalization of the ISR data is overestimated by factor two.  This was observed in  $pp$ collisions in the NA49 experiment \cite{na49}, and in $ep$ \cite{zeus} collisions.
Here  we concentrate on the spin dependence of neutron production in the fragmentation region of the polarized beam caused by the absorptive corrections.

\section{Absorptive corrections to the pion pole}

Pion exchange contribution to the amplitude of neutron production, $pp\to nX$ in
Born approximation has the form
 \beq
A^B_{p\to n}(\vec q,z)=\frac{1}{\sqrt{z}}\,
\bar\xi_n\left[\sigma_3\,\tilde q_L+
\vec\sigma\cdot\vec q_T\right]\xi_p\,
\phi^B(q_T,z)\,,
\label{100}
 \eeq
 where $\vec\sigma$ are Pauli matrices;  $\xi_{p,n}$ are the proton or
neutron spinors;  $\vec q_T$ is the transverse component of the momentum transfer;
$\tilde q_L=(1-z)\,m_N$.  Fractional momentum $z$ is related to the invariant mass $M_X$ of $X$, 
$z\approx 1-M_X^2/s$ for $1-z\ll1$, where $\sqrt{s}$ is the c.m. energy of $pp$ 
collision.  The 4-momentum
transfer squared $t$ has the form,
$t=-(\tilde q_L^2+q_T^2)/z$.

\begin{figure}[htb]
\centerline{
 {\includegraphics[height=7cm]{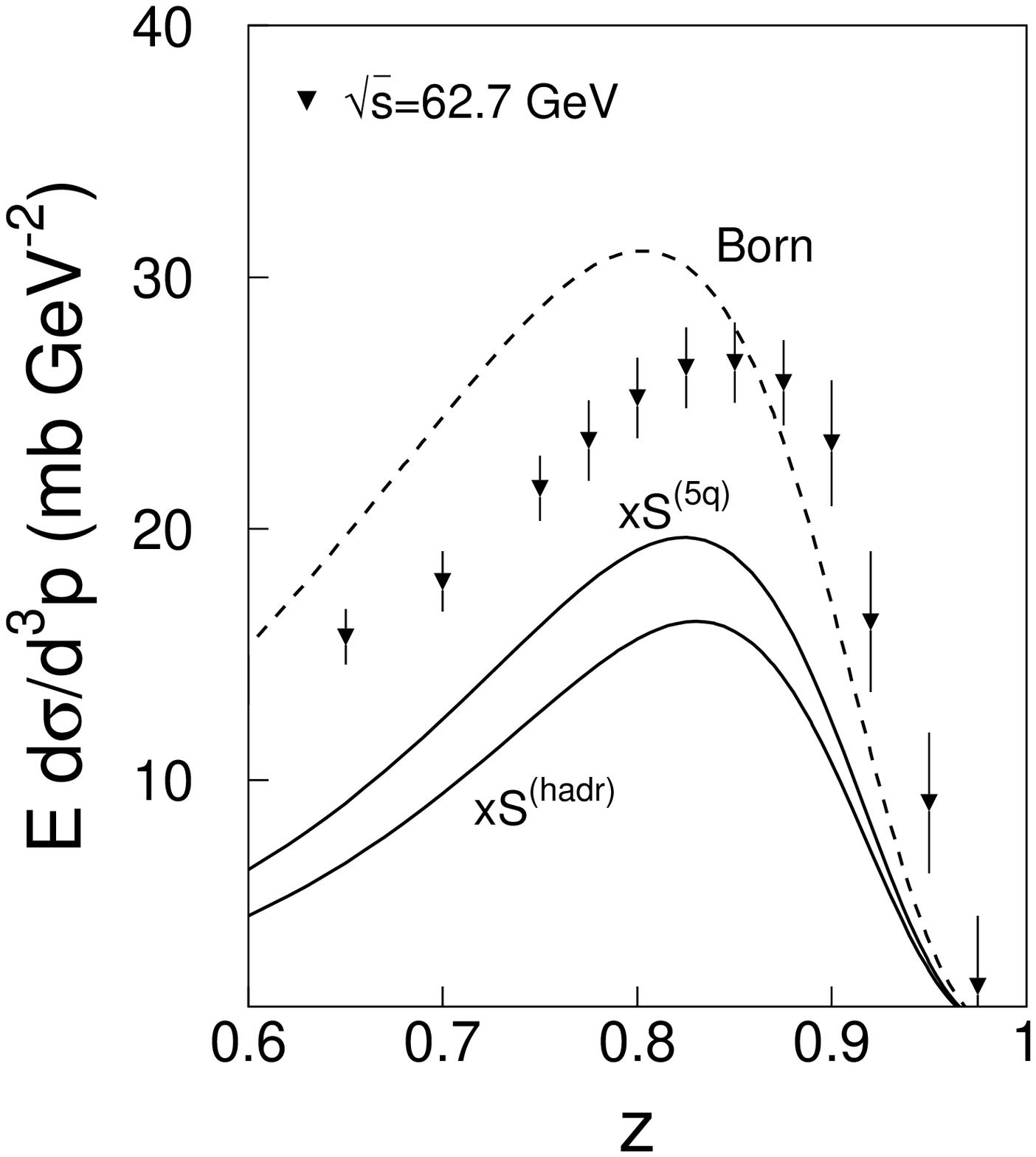}}
{\includegraphics[height=7cm]{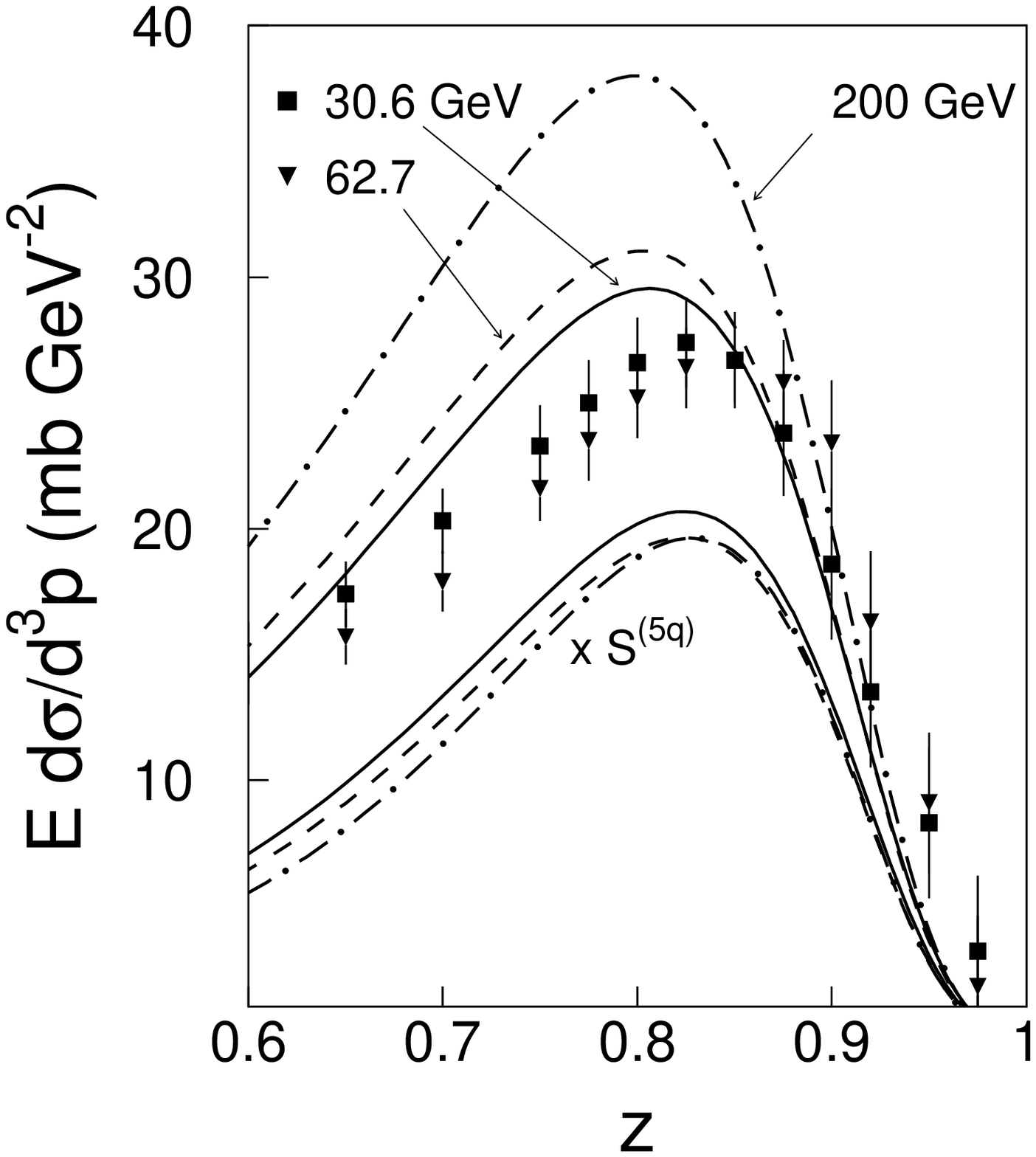}}
 }
\caption{\label{fig:isr}
 {\it Left panel: } Born approximation (dashed curve) for neutron
production and ISR data \cite{isr} at $\sqrt{s}=60.7\GeV$ and $p_T=0$. Two
solid curves upper and bottom show the effect of absorptive corrections
calculated in the dipole approach ($\times S^{(5q)}$) and in hadronic
representation ($\times S^{(hadr)}$) respectively.
  {\it Right panel:} Energy dependence of inclusive neutron production. Three
upper curves present the forward cross section at $\sqrt{s}=30.6\GeV$
(solid), $62.7\GeV$ (dashed) and $200\GeV$ (dotted-dashed) calculated in the
Born approximation. These curves corrected for absorption ($\times
S^{(5q)}$) are presented by three curves at the bottom. Data at
$\sqrt{s}=30.6\GeV$ and $62.7\GeV$ \cite{isr} are depicted by squares and
triangles respectively
 }
 \end{figure}

The differential cross section of inclusive
neutron production reads \cite{bishari,2klp,kpss},
 \beqn
z\,\frac{d\sigma^B_{p\to n}}{dz\,dq_T^2}&=&{1\over s}
\left|A^B_{p\to n}(\vec q_T,z)\right|^2=
\nonumber\\ &=&
\left(\frac{\alpha_\pi^\prime}{8}\right)^2
|t|G_{\pi^+pn}^2(t)\left|\eta_\pi(t)\right|^2
(1-z)^{1-2\alpha_\pi(t)}
\sigma^{\pi^+ p}_{tot}(s'=M_X^2),
\label{146}
 \eeqn
where $\eta_\pi(t)$ is the signature factor, and $G_{\pi^+pn}^2(t)=\exp(t\,R_1^2)$ is the pionic formfactor (see details in \cite{kpss}).  The results of calculation with the Born approximation, Eq.~(\ref{146}), at 
$\sqrt{s}=200,\ 62.7$ and $30.6\GeV$ are
depicted versus ISR data \cite{isr} in Fig.~\ref{fig:isr}.
 The data are shown at two energies $\sqrt{s}=30.6\GeV$ and $62.7\GeV$.
Correspondingly, we used these energies in our calculations. One can see
that the Born approximation considerably exceeds the data and leads to a wrong energy dependence.

Absorptive corrections, or initial/final state interactions look simpler and factorize in impact parameters. Therefore,
we switch to a Fourier transformed amplitude (\ref{100}) which has the form,
 \beq
f^B_{p\to n}(\vec b,z)=\frac{1}{\sqrt{z}}\,
\bar\xi_n\left[\sigma_3\,\tilde q_L\,\theta^B_0(b,z)-
i\,\frac{\vec\sigma\cdot\vec b}{b}\,
\theta^B_s(b,z)\right]\xi_p\,.
\label{150}
 \eeq
 Here
\beqn
\theta^B_0(b,z) &=& \int d^2q\,e^{i\vec b\vec q}\,
\phi^B(q_T,z)
\nonumber\\ &=&
N(z)\left\{i\,\frac{\pi\alpha_\pi^\prime}{2z\beta^2}\,
K_0(b/\beta) +
\frac{1}{1-\beta^2\epsilon^2}\,
\left[K_0(\epsilon b)-K_0(b/\beta)\right]\right\}\,;
\label{154}
 \eeqn
 \beqn
\theta^B_s(b,z) &=& {1\over b}
\int d^2q\,e^{i\vec b\vec q}\,
(\vec b\cdot\vec q)\,\phi^B(q_T,z)
\nonumber\\ &=&
N(z)\left\{i\,\frac{\pi\alpha_\pi^\prime}{2z\beta^3}\,
K_1(b/\beta) +    
\frac{1}{1-\beta^2\epsilon^2}\,
\left[\epsilon\,K_1(\epsilon b)-\frac{1}{\beta}\,K_1(b/\beta)
\right]\right\}\,;
\label{164}
 \eeqn
and
 \beqn
N(z) &=&\frac{1}{2}\,g_{\pi^+pn}\,
z(1-z)^{\alpha^\prime_\pi(m_\pi^2+q_L^2/z)}
e^{-R_1^2 q_L^2/z}
A_{\pi p\to X}(M_X^2)\,
\nonumber\\
\epsilon^2&=&q_L^2+z m_\pi^2\,,
\nonumber\\
\beta^2&=&{1\over z}\,\left[
R_1^2-\alpha_\pi^\prime\,\ln(1-z)\right]\,.
\label{166}
 \eeqn

The process under consideration at large $z\to1$ is associated with
creation of a large rapidity gap (LRG), $\Delta y=|\ln(1-z)|$, where no
particle is produced. Absorptive corrections may be also interpreted as a
suppression related to the survival probability of LRG. 

\begin{figure}[htb]
\centerline{
 {\includegraphics[height=3cm]{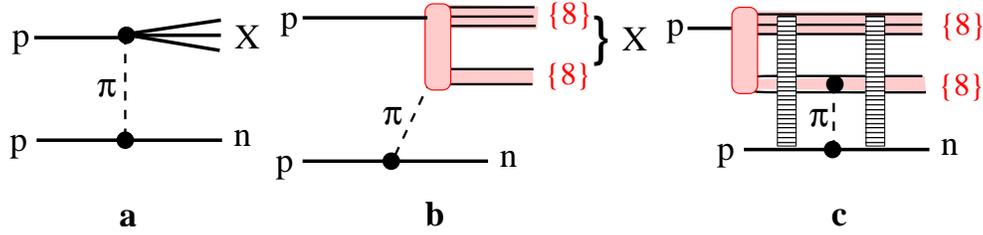}}
 }
\caption{\label{fock-pp}
 {\bf a:} Born graph with single pion exchange and excitation of the projectile 
proton, $p+\pi\to X$;
 {\bf b:} inelastic proton-pion interaction interaction, $p+\pi\to X$, via
color exchange leading to production of two color-octet dipoles which
hadronize further to $X$;
 {\bf c:} Fock state representation of the previous mechanism. A color 
octet-octet dipole which is a 5-quark Fock component of the projectile proton, 
interacts with the target proton via 
$\pi^+$ exchange. This 5-quark state may experience initial and final state 
interaction via vacuum quantum number (Pomeron) exchange with the nucleons 
(ladder-like strips). }
 \end{figure}

To identify the projectile Fock states responsible for absorptive corrections we start with
Fig.~\ref{fock-pp}a containing the amplitude of pion-proton inelastic
collision $\pi+p\to X$. The latter is usually described as color exchange
leading to creation of two color octet states with a large rapidity
interval $\sim \ln(M_X^2/s_0)$ ($s_0=1\GeV^2$) as is illustrated in
Fig.~\ref{fock-pp}b. 
 Perturbatively, the interaction is mediated by gluonic exchanges.
Nonperturbatively, e.g. in the string model, hadron collision looks like crossing
and flip of the strings. 

According to Fig.~\ref{fock-pp}b one may think that this is the produced color
octet-octet state which experiences final state interactions with
the recoil neutron. On the other hand, at high energies multiple 
interactions become coherent, and both initial and final state 
interactions must be included. This leads to a specific space-time 
development of the process at high
energies, namely, the incoming proton fluctuates into a 5-quark state
$|\{3q\}_8\{\bar qq\}_8\ra$ long in advance of the interaction with the target via pion exchange, as is illustrated in Fig.~\ref{fock-pp}c. 

This picture can be also interpreted in terms of the Reggeon calculus \cite{kpss},
which helps to identify what was overlooked in the calculation of absorptive corrections done in \cite{3n}.

The resulting amplitude exposes both initial and final state attenuation of the 5-quark state,
 \beq
f_{p\to n}(b,z)=f^B_{p\to n}(b,z)\,S(b,z),
\label{195}
 \eeq
which are incorporated via the survival amplitude $S(b,z)$. 

We evaluate the survival amplitude $S(b,z)$ within two models based on the color-dipole and hadronic representations.

\begin{figure}[htb]
\centerline{
 {\includegraphics[height=6.5cm]{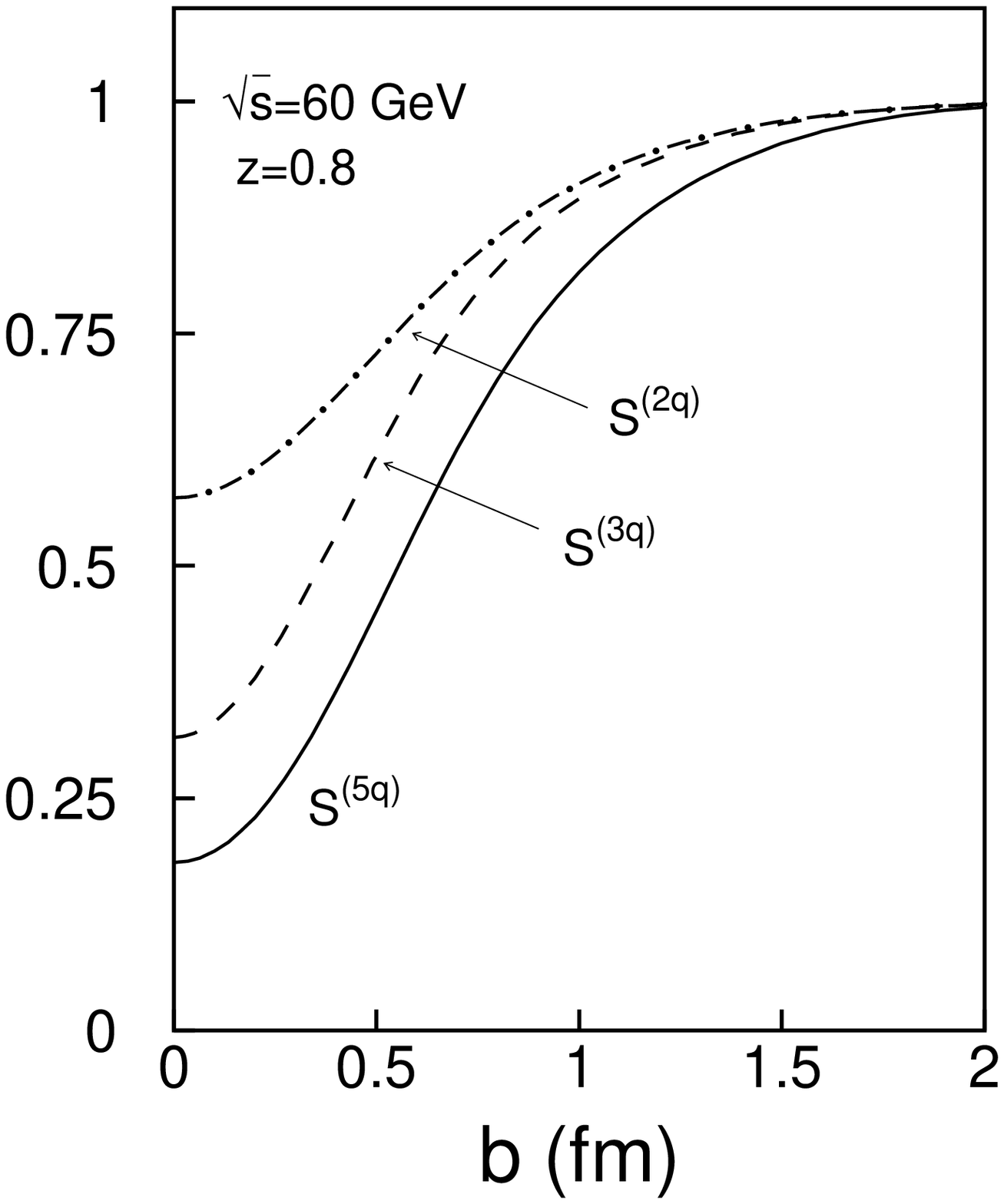}}
 {\includegraphics[height=6.5cm]{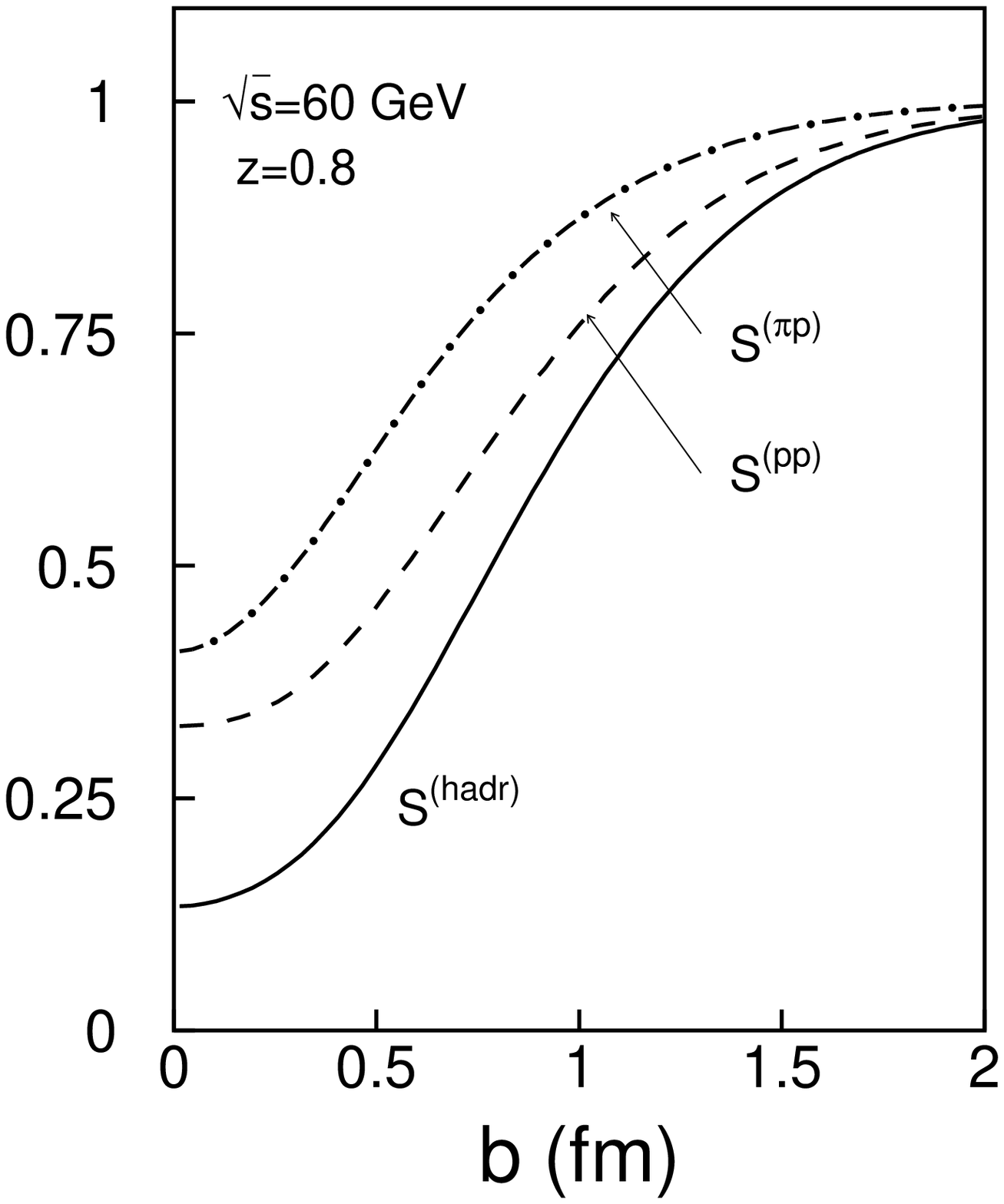}}
}
 \caption{ Partial survival amplitude $S(b,z)$ at $\sqrt{s}=60\GeV$ and
$z=0.8$. {\it Left panel:} Survival amplitudes $S^{(2q})(b,z)$  and $S^{(3q})(b,z)$ depicted by
dot-dashed and dashed curves, respectively. Solid curve corresponds to $S^{(5q})(b,z)$.
{\it Right panel:} Survival amplitudes evaluated in hadronic representation. }
 \label{survival}
 \end{figure}

{\it Color-dipole model.}\\
 According to the dual parton model \cite{capella} one can present the survival amplitude of a 5-quark state with an accuracy $1/N_c^2$ as,
 \beq
S^{(5q)}(b) = S^{(3q)}(b)\,S^{(q\bar q)}(b)
=\left[1-\Im\Gamma^{(3q)p}(b)\right]
\left[1-\Im\Gamma^{(\bar qq)p}(b)\right]\,.
\label{290}
 \eeq
 similar to Eq.~(\ref{240}),  
The elastic amplitudes $\Gamma^{(\bar33)p}(b)$ of a color $\{\bar 33\}$ 
dipole interacting with
a proton can be calculated in terms of  the partial dipole elastic amplitudes, for which
a saturated model was proposed recently \cite{krs1,kpss}.
The results for $S^{(3q)}(b)$, $S^{(q\bar q)}(b)$ and their product $S^{(5q)}(b)$ at $\sqrt{s}=60\GeV$ and $z=0.8$ are depicted in
Fig.~\ref{survival} (left panel).

{\it Hadronic representation.}\\
One can expand the 5-quark Fock state over the hadronic basis,
 \beq
\left|\{3q\}_8\{\bar qq\}_8\right\ra =
d_0|p\ra + d_1|N\pi\ra + d_2|N2\pi\ra + ...\,.
\label{200}
 \eeq
Since the admixture of sea quarks in the proton is small, the
projection of the 5-quark state to the proton must be small too. 
It is natural to assume that the amplitude
$d_1$ is the dominant one, since both states $|\{3q\}_8\{\bar
qq\}_8\ra$ and $|N\pi\ra$ have the same valence quark content.
Correspondingly, the absorption corrected partial amplitude gets the form,
 \beq
f_{p\to n}(b,z)=f^B_{p\to n}(b,z)\,S^{(hadr)}(b)\,,
\label{220}
 \eeq
 where 
 \beq
S^{(hadr)}(b) = S^{\pi p}(b)\,S^{pp}(b)
=\left[1-\Im\Gamma^{pp}(b)\right]
\left[1-\Im\Gamma^{\pi p}(b)\right]\,.
\label{240}
 \eeq
The elastic partial amplitudes $\Im\Gamma^{hp}(b)$ were extracted in a model independent way directly from data on elastic scattering \cite{k3p,kpss}.

 Our results for $S^{(hadr)}(b,z)$ are depicted in Fig.~\ref{survival}, right panel. 
 The survival amplitude suppression factor $S^{(hadr)}(B,z)$ is rather similar to one calculated in the dipole model, $S^{(51)}(b,z)$.

\section{Absorption corrected spin amplitudes}

\begin{figure}[htb]
\centerline{
  \scalebox{0.4}{\includegraphics{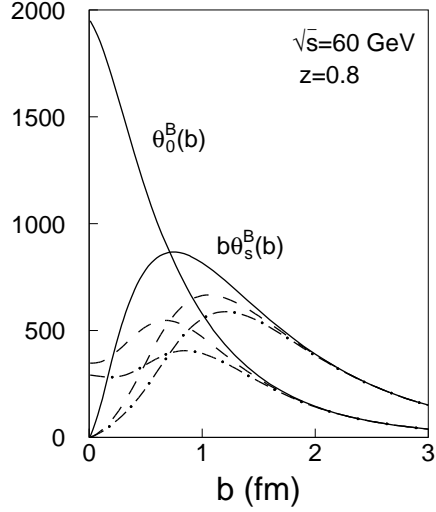}}
}
 \caption{Partial spin amplitudes for neutron production.
Solid curves show
the result of Born approximation. Dashed and dot-dashed curves include
absorptive corrections calculated in the dipole approach ($\times
S^{(5q)}(b,z)$) and in hadronic model ($\times S^{(hadr)}(b,z)$),
respectively.}
 \label{spin-amplitudes}
 \end{figure}

 An example of absorption corrected impact-parameter-dependent spin amplitudes is shown in Fig.~\ref{spin-amplitudes} for $\sqrt{s}=60\GeV$ and $z=0.8$ in comparison with the Born approximation.

 Now, it is straightforward to Fourier transform the absorption corrected  amplitudes (\ref{195}) back to momentum representation,
 \beq
A_{p\to n}(\vec q,z)=\frac{1}{\sqrt{z}}
\bar\xi_n\left[\sigma_3 \tilde q_L\,\phi_0(q_T,z)+
\vec\sigma\cdot\vec q_T\phi_s(q_T,z)\right]\xi_p,
\label{520}
 \eeq
 where 
   \beqn
\Re\phi_0(q_T,z)&=&\frac{N(z)}{2\pi(1-\beta^2\epsilon^2)}
\int\limits_0^\infty db\,b\,J_0(bq_T)
\left[K_0(\epsilon b)-K_0\left({b\over\beta}\right)\right]
S(b,z)\,;
\nonumber\\
\Im\phi_0(q_T,z)&=&\frac{\alpha_\pi^\prime N(z)}{4z\beta^2}
\int\limits_0^\infty db\,b\,J_0(bq_T)\,
K_0\left({b\over\beta}\right)\,
S(b,z)\,;
\label{550}
 \\
 q_T\Re\phi_s(q_T,z)&=&\frac{N(z)}{2\pi(1-\beta^2\epsilon^2)}  
\int\limits_0^\infty db\,b\,J_1(bq_T)
\left[\epsilon\, K_1(\epsilon b)-
{1\over\beta}\,K_1\left({b\over\beta}\right)\right]   
S(b,z)\,;
\nonumber\\
q_T\Im\phi_0(q_T,z)&=&\frac{\alpha_\pi^\prime N(z)}{4z\beta^3}
\int\limits_0^\infty db\,b\,J_1(bq_T)\,
K_1\left({b\over\beta}\right)\,   
S(b,z)\,.
\label{570}
 \eeqn
 
 The forward neutron production cross section corrected for absorption is
compared with the ISR data \cite{isr} in Fig.~\ref{fig:isr}. The two models for
absorption, dipole and hadronic, are presented on the left panel by solid curves, 
upper and bottom respectively. The results of both models are pretty 
close to each other, but substantially underestimate the ISR data.
  It was argued, however, in \cite{kpss} that the normalization of these data is twice overestimated. This is supported by recent direct measurements of $pp\to nX$ by the NA49 experiment \cite{na49} and by comparison with neutron production in DIS \cite{zeus} (see discussion in \cite{kpss}).
Otherwise, the shape of the $z$-dependence is reproduced pretty well.
Also the energy dependence of the cross section is much improved after inclusion of absorption. Apparently, the 
steep rise of the cross section with energy observed in Born approximation
is nearly compensated by the falling energy dependence of LRG survival 
amplitudes.
 The results for $q_T$ dependent differential cross section can be found in \cite{kpss}.

\begin{figure}[htb]
\centerline{
 {\includegraphics[height=5.5cm]{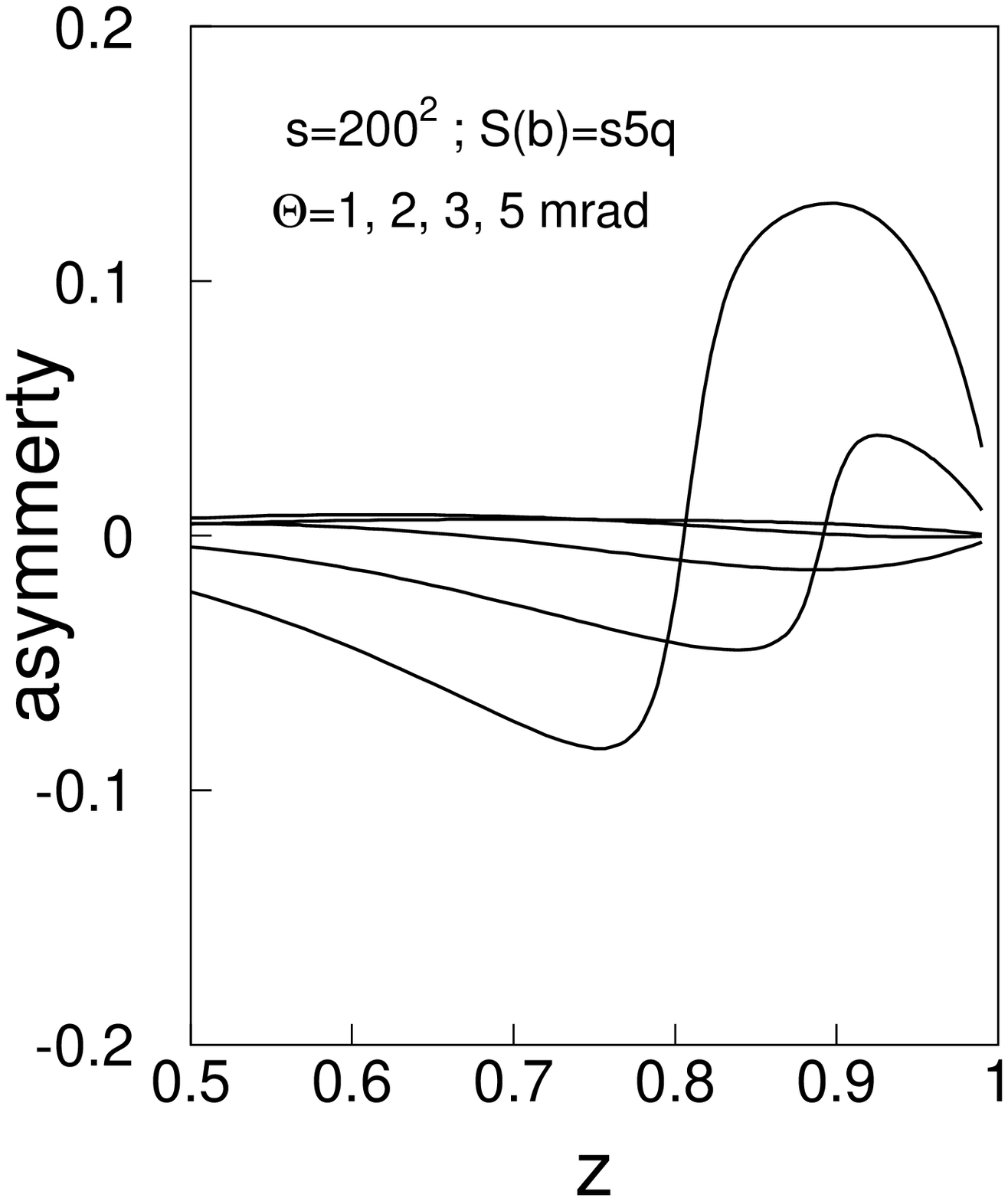}}
 {\includegraphics[height=5.5cm]{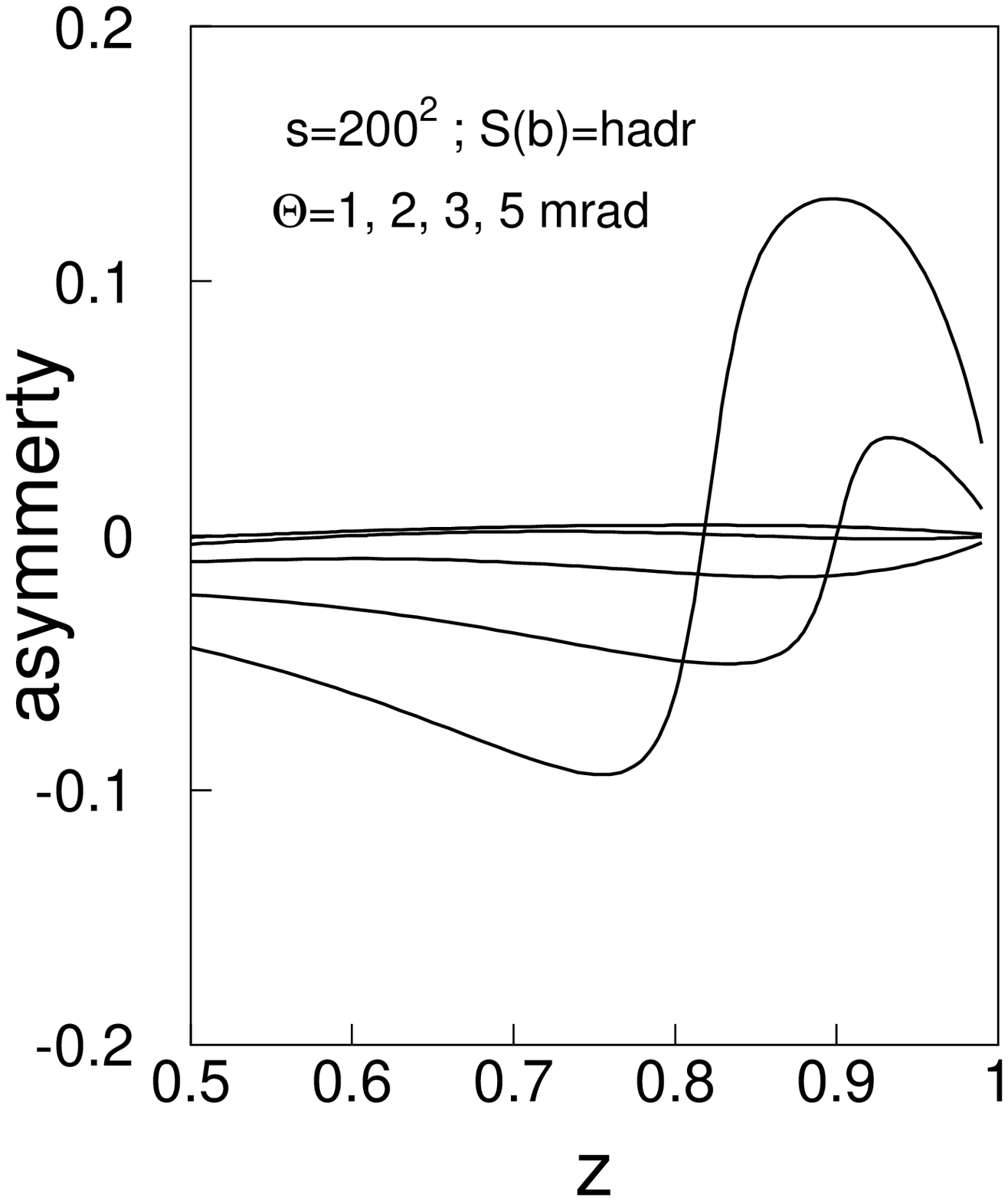}}
 {\includegraphics[height=5.5cm]{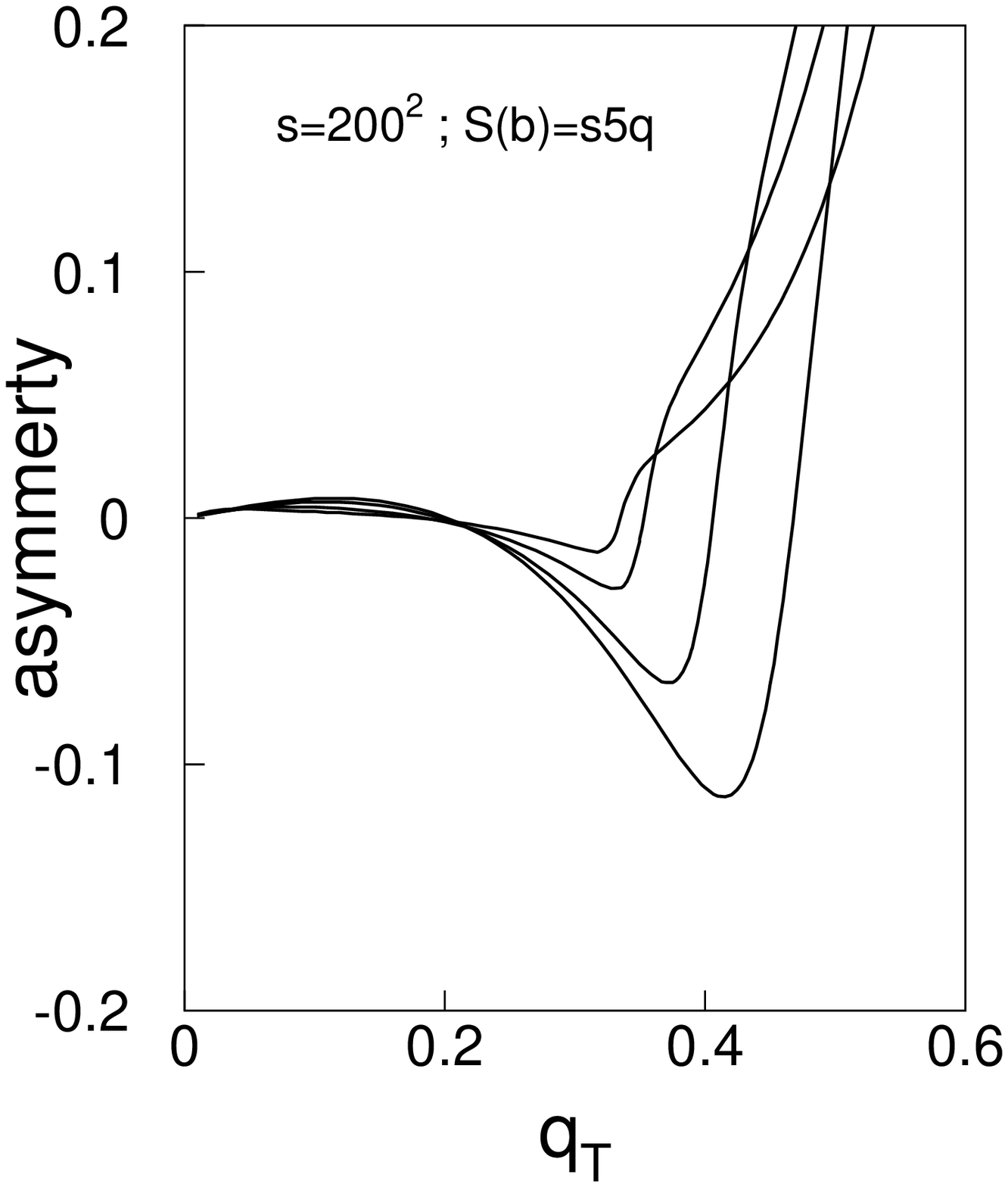}}
}
 \caption{The curves in the first and second left panels correspond to the absorption corrections calculated in the dipole and hadronic models respectively. The spin asymmetry is calculated at 
five different fixed angles between the neutron and initial proton 
momenta in the c.m. of collision $\theta_{c.m.}=0.001 - 0.005$. 
Correspondingly, neutron transverse and longitudinal momenta correlate,
$q_T=\theta_{c.m.}\,z\,\sqrt{s}/2$. 
The last panel shows the $q_T$ dependence
of $A_N$ calculated in the dipole model at four fixed values of $z=0.6-0.9$.}
 \label{asymmetry}
 \end{figure}

As far as the spin amplitudes are known, one can calculate the single spin asymmetry,
 \beq
A_N(q_T,z)=\frac{2q_Tq_L\phi_0(q_T,z)\phi_s(q_T,z)}
{q_L^2\left|\phi_0(q_T,z)\right|^2 +
q_T^2\left|\phi_s(q_T,z)\right|^2}\,
\sin(\delta_0-\delta_s)\,,
\label{620}
 \eeq
 where
 \beq
\tan\delta_{0,s}=\frac{\Im\phi_{0,s}(q_T,z)}
{\Re\phi_{0,s}(q_T,z)}\,.
\label{640}
 \eeq
 Our predictions for  $A_N(q_T,z)$ are presented in Fig.~\ref{asymmetry}. 
Notice that the spin asymmetry  depicted in the two left panels, demonstrate practically no model dependence of the spin asymmetry.

Unfortunately, our calculations at transverse momenta as small as was measured at RHIC \cite{phenix} considerably underestimate the data. This is not a surprise, $q_T^2\sim 0.01 GeV^2$ is so small, that hadronic spin effects in other reactions are usually vanishingly small. 
The spin asymmetry which arises due to pion pole with absorptive corrections, develops
a double zero at $q_T\to0$. This is because both, the spin-flip amplitude and the relative phase, Eq.~(\ref{640}), vanish.
However,  at larger transverse momenta the single spin asymmetry reaches an appreciable value. One should include other mechanisms, for instance interference of pion and $a_1$ exchanges, which is enhanced by a proper phase shift, $\pi/2$ and energy independence of the diffractive amplitude $\pi p \to a_1 p$.
 
\begin{theacknowledgments}
 This work was supported in part by Fondecyt (Chile) grants 1050519 and 1050589, and by DFG (Germany)  grant PI182/3-1.

\end{theacknowledgments}

\bibliographystyle{aipproc}

\end{document}